\documentclass[twocolumn,twoside,slac]{revtex4}
\usepackage{graphicx}
\usepackage{fancyhdr}
\begin{document}

\title{A Modular Object Oriented Data Acquisition System for the Gravitational Wave AURIGA Experiment}


\author{A. Ceseracciu}
\affiliation{Universit\'{a} di Padova / Laboratori Nazionali di Legnaro - INFN}
\author{G. Vedovato}
\affiliation{Laboratori Nazionali di Legnaro - INFN}
\author{A. Ortolan}
\affiliation{Laboratori Nazionali di Legnaro - INFN}

\begin{abstract}
The new Data Acquisition system for the gravitational wave detector AURIGA has been designed from the ground up in order to take advantage of hardware and software platforms that became available in recent years; namely, i386 computers running Linux-based free software.\\ 
This paper describes how advanced software development technologies, such as Object Oriented design and programming and CORBA infrastructure, were exploited to realize a robust, distributed, flexible, and extensible system.\\ 
Every agent of the Data Acquisition System runs inside an heavyweight framework, conceived to transparently take care of all the agents intercommunication, synchronization, dataflow. It also presents an unified interface to the command and monitoring tools. The DAQ logic is entirely contained in each agent's specialized code. In the case of AURIGA the dataflow is designed as a three tier: frontend, builder, consumer. Each tier is represented by a set of agents possibly running on different hosts.\\
This system is well fit for projects on scales comparable to the AURIGA experiment: permanent and temporary data storage is based on the Frame format adopted by the gravitational wave community, and the design is reliable and fault-tolerant for low rate systems.
\end{abstract}

\maketitle

\thispagestyle{fancy}


\section{\label{hardware}THE AURIGA DATA ACQUISITION SYSTEM}

\subsection{The AURIGA detector}
AURIGA is a resonant gravitational wave (gw) detector located at the INFN-
National Laboratories of Legnaro in Italy, designed to detect  short  bursts
of gravitational radiation at characteristic frequencies of about  1  $kHz$
within a $50\div 80 \ Hz$ bandwidth. The goal of gravitational wave experiments is  to
make  astronomical  observations  of  gw   emitted   during   catastrophic
astrophysical phenomena such  as   supernova  explosions  or  a  black  hole
collisions.\\
AURIGA is a member of a network of gw  detectors,
either  interferometers  or  resonant  cryogenic  bars.
Operating gw detectors in a tight coordinated network
allows for a drastic reduction of spurious signals and an experimental
determination of the false alarm rate. In addition, networks of 3, or
preferably more, detectors ensure the complete reconstruction of a gw
event, the measure of its velocity of propagation, and the recognition of
the Riemann tensor intrinsic signatures.
The network task poses a first  demanding
request to the data acquisition and analysis system of the AURIGA  detector,
namely an accurate synchronization with the GPS time
at least  within  few $\mu$sec.\\
Other  requirements  include:  the  capability  of
handling a large amount of data ($\sim 3 \ GB$ per  day),  including  data  from
the auxiliary channels which monitor the  detector  environment;  continuous
data collecting, 24 hours a day, for many years; 
real time data conditioning and data
preprocessing, for the purpose of monitoring and diagnosing
the detector performance and to provide
data integral to gravitational wave analysis.\\
The preprocessed data has to be analyzed off-line  searching  for millisecond gw bursts from Supernova explosions, periodic  signal  from
galactic pulsars and cross correlated with the  output  of  other  detectors
searching for the stochastic gw background.

\subsection{The DAQ hardware setup}
\begin{figure*}[t]
\centering
\includegraphics[width=135mm]{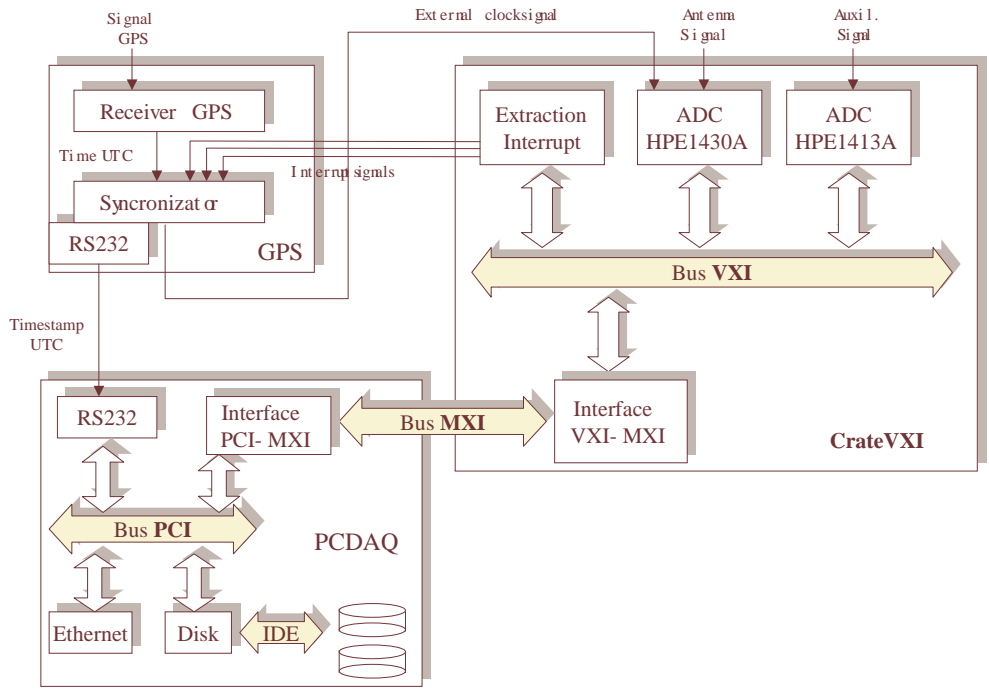}
\caption{Main hardware components of the DAQ system, and dataflow between them.} \label{daqfig}
\end{figure*}

The AURIGA DAQ system acquires  and  archives  
the signal channels devised for gw
measurements and for controlling,
monitoring and diagnosing the detector and its environment.\\
In fig. \ref{daqfig} we report  a  schematic
diagram of this system. The dc SQUID output  is  sampled  at  4882.8125  Hz
with a 23 bit (19 effective) AD converter (HP  E1430A)  housed  into  a  VXI
crate (VXI is an industrial standard bus  for  electronic  instrumentation).
The  data  from  the  accessory  instrumentation,  such  as  accelerometers,
seismometers, electromagnetic probes etc. are sampled  at  rates  between  1
and 200 Hz with a 32 multiplexed channels, 16 bit, AD  converter  HP  E1413A
housed in the same VXI crate.

\subsection{1 $\mu$s time synchronization}
The HP E1430A stores the sampled data in its 8 $MB$ FIFO memory that has been divided into 64 $kB$ blocks. When a data block is ready, an interrupt signal (IRQ) is generated in the VXI bus.  The IRQ signal notifies to a process in the acquisition workstation that a data block is ready and the data are read out.  The same IRQ is sent to the GPS synchronization apparatus to date the read data block. The IRQ generation mechanism has an intrinsic jitter of $\cong 0.1 \mu sec$ while the IRQ propagation lines introduce a fixed delay in the time associated to each data block of $1104.4 \pm 0.4 \mu sec$.

\section{THE PREVIOUS DATA ACQUISITION SOFTWARE}

\subsection{Technology: integrated and proprietary}
To design the previous data acquisition and data analysis software, the CASE (Computer Aided Software Engineering) Artifex 4.2 was chosen. This tool has been used to translate the DAQ and on-line data analysis models into software to control distributed multi-process applications (see \cite{olddaq}).\\  
Artifex tools provide the user with an object based decomposition of the software architecture with intuitive graphical interfaces, the software modeling based on the Petri Network, a tool for the software documentation (HTML or LaTeX) and a tool for multiplatform code generation  (Windows NT or Unix: Solaris 2.6, Digital Unix 4.0).

\subsection{Development process and limitation}
Being the first attempt at building an acquisition system for the experiment, its evolution was necessarily driven by continuously changing requirements. The CASE system helped coping with the added complexity of the new tasks, up to a point where the modeling tools it provided became very difficult to manage and understand because of the sheer complexity needed to address some issues within the somewhat limited framework provided.\\
Even though the performance was satisfying and reliability good, this system eventually seemed too constrictive for the future evolution of the requirements.

\section{THE NEW DATA ACQUISITION SOFTWARE}

\subsection{Evolution of requirements}
The first requirement for the new DAQ is: provide all the basic functionality offered by the old system.\\
The performance required to handle the readout of the AURIGA detector is low, when compared to modern hardware limits. The net throughput of the main channel, as introduced in section \ref{hardware}, is 32 bits at 5 $kHz$, or 20 $kB/sec$, while the auxiliary channels yield about 10 $kHz$, depending on the configuration.
The main requirement for the DAQ system is to guarantee integrity of the data taken during the long acquisition runs of AURIGA.\\
The amount of data produced per day is about 1.7 $Gb$, or 2.5 $Gb$ including auxiliaries. Modern hard drives easily provide this capacity; nevertheless, since a single run could last several months, it is necessary to allow to change the 
writing device without blocking the acquisition process.\\
The time synchronization with respect to the UTC is required to be accurate to 1 $us$. This accuracy is obtained by custom acquisition hardware. The software is responsible to recognize and recover hardware problems or glitches. This hardware-based solution allows the software to run on a non-realtime OS such as the standard Linux kernel.\\
In acquisition operations is often necessary the quickest intervention time in case of problems. To allow that, the acquisition software must provide a logging infrastructure supporting multiple real-time terminals, and association of custom alarms to some events.\\
Each acquisition process handles a data flow coming from different ADCs, and bearing a well defined physical meaning. A necessary diagnostic tool is hence a data monitor, able to attach in real-time to any process that owns a data queue, and to display the data both in the time and in the frequency domain.\\
The physics analysis system uses the same architecture as the acquisition software for command interfaces, process intercommunication, I/O. The DAQ software design must hence consider these needs.

\subsection{Technology gap}
%%sw components
The programming language chosen for the implementation of the data acquisition system is C++.\\
The main advantage of the C++ language is support for object-oriented programming paradigm. This allows to formally structure the code by logic rather than functional criteria, to enforce strict type checking, and to manage dependencies between different code sections. For large-sized projects these features make C++ better suited than C.\\
On the other hand, the current most popular OO language alternative to C++, Java, is unsuited for this project because of the following two reasons: i) a direct hardware interface, needed e.g. to handle the ADC registers, is not available from inside a virtual machine like Java's one; and ii) the physics analysis routines, essentially dealing with digital signal processing, need number crunching performance that Java can not provide.\\
Process intercommunication is built over a CORBA middleware. The main benefit of this technology is the transparent interface over TCP/IP and the strong typed interface for remote methods.\\
The I/O format is the \em IGWD frame\em \cite{framelib}, established by the collaboration between the leading interferometric experiments VIRGO \cite{virgo} and LIGO \cite{ligo}. This format makes easier the data exchange between different experiments, by giving a structured container for gravitational wave data. The space overhead is negligible, amounting to less than 1\%{ } for the typical AURIGA data files.

\subsection{Development tools and components}
The C++ compiler is the well known GCC. This compiler proved good compliance to advanced C++ features and good code efficiency.\\
The OS is GNU/Linux. This choice was motivated by the established stability of this operating system and the great choice of development tools, and by the good performance on the \em i386\em, the most convenient platform currently available on the market, with respect to the performance/price ratio.\\
The chosen Linux distribution was Red Hat 6.2, stable though obsolete, and the adopted IDE (Integrated Development
Environment) was KDevelop \cite{kdev}.\\
The Red Hat choice was due mainly because of the huge share that it has in the Linux installation base. This figure is especially important due to the \em open source \em nature of this software.\\
The choice of an obsolete version of the distribution is due to a compatibility requirement for the \verb ni-vxi \ driver provided by National Instruments, the available version of which was not able to work with the 2.4 family of the Linux kernel. Nowadays, National Instruments announces the availability
of an updated version of this component that allows for
a migration to more recent Linux distributions (e.g. Red Hat 7.3).\\
The chosen CORBA 2.3 \cite{corba} implementation was omniORB v3.0.4 \cite{omnidoc}, released as open-source by AT\&T. This implementation looks like the most convenient choice because of its stability, completeness and full multithread architecture.\\
File access is done via the \em frame library \em \cite{framelib}, developed at VIRGO \cite{virgo}. As this library is programmed in C, a minimal C++ wrapper was made to use it from the acquisition system.\\
%%manca: VEGA,ROOT,QT........

\subsection{\label{opensource}Developing on  Open Source Software}
The migration to a fully open-source development environment follows well-established trend over the past few years in scientific computing.\\
The actual motivation was to exploit the performance/cost efficiency attained by the $i386$ platform. This is yet another demonstration of the boosting effect that cheap powerful hardware had in the open-source revolution.\\
The first challenge in going open-source has been the choice of tools. While some choices were pretty easy because of their widespread diffusion, like the Linux OS kernel and the GCC compiler. Other choices were more controversial, e.g. the CORBA implementation, the graphical interface toolkit, some HEP specific components. Some of these choices were actually revised and changed during the development, and some components underwent major management changes, namely the OmniORB package. The process of adopting an alternative component has never been too cumbersome. The pervasive agreement on interfaces that makes possible the very development of any complex open-source software actually allowed to produce a modular system, and indirectly helped to design our custom code in a modular way. This is traditionally much more difficult when developing on proprietary integrated software.\\
Concluding, notwithstanding the steeper learning curve, open-source software gave a better insight and allowed to focus on design rather than working around the development environment limitations.

\subsection{Agent-wise system architecture}
\begin{figure}
\includegraphics[width=80mm]{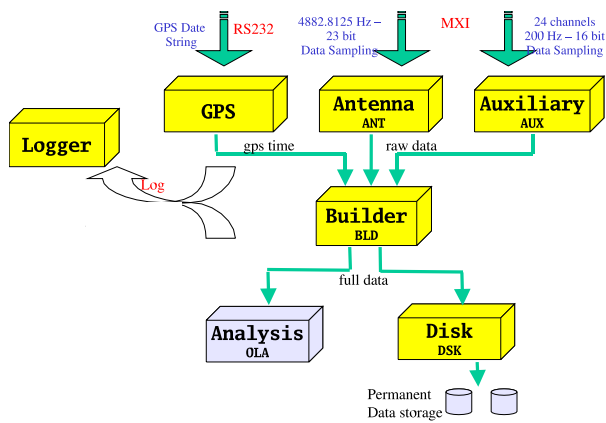}
\caption{The process-wise architecture of the DAQ software. Displays the three layered structure and the central role of the BLD process.} \label{archfig}
\end{figure}

Agent-wise, the acquisition system can be seen as a set of processes organized as cascading boxes, and specialized for different tasks. This architecture can be described as an intense client-server: each element of the cascade is indeed a client for the previous one, and a server to the next one. This definition is even more appropriate when considering the polling interface for the data flow.\\
The actual processes' architecture and dataflow is depicted in figure \ref{archfig}.\\
%%TODO: describe figure...

\section{THE FRAMEWORK: PCL (Process Control Library)}
The PCL is packaged as a shared library. The PCL makes extensive use of advanced C++ features: simple and multiple inheritance, abstract classes, exception handling, template programming.\\
All the agents of the acquisition system, with the exception of pure clients, use the PCL framework. They are hence called \em PCL processes\em.\\
As soon as a PCL process is started, it makes its remote invocation interface by registering itself to a CORBA nameserver. This remote interface allows clients to communicate with the PCL process.\\
Production of data takes place inside an asynchronous thread. The synchronization of the production with the client is done by means of special queues.\\
The PCL framework provides a template for processes to implement their custom code. A graphical presentation is provided in figure \ref{processfig}, and a detailed discussion is later in this section.

\subsection{\label{queues}Queues}
\begin{figure*}[t]
\centering
\includegraphics[width=135mm]{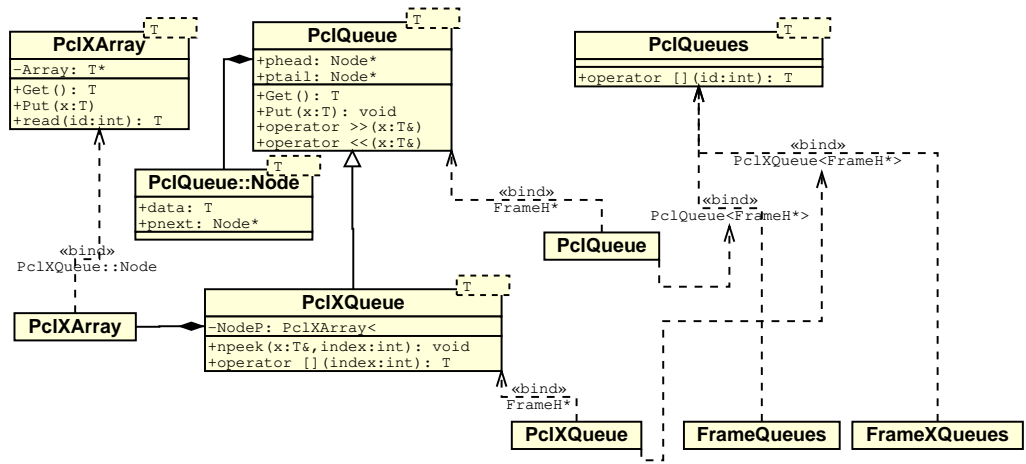}
\caption{Inheritance relationship between the transient queue types. The elementary item is a FrameH* frame} \label{queuefig}
\end{figure*}

\begin{figure*}[t]
\centering
\includegraphics[width=135mm]{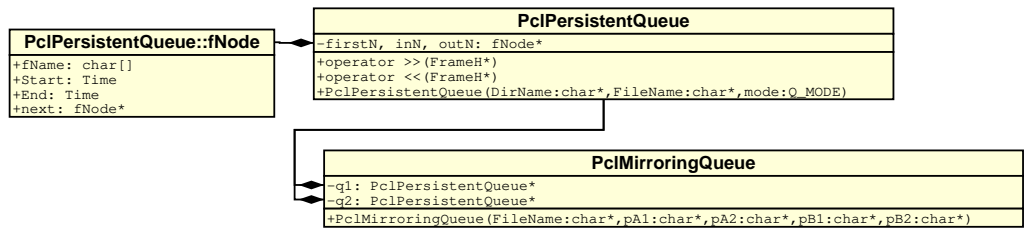}
\caption{Inheritance relationship between the persistent queue types.} \label{persqueuefig}
\end{figure*}

In every PCL process temporary and permanent data storage, communication, data monitoring, depend on some queue. Queues inside the acquisition system are thus remarkably important and specialized.\\
Three fundamental types of queues are provided by the PCL: \verb PclQueue \ and the derived \verb PclXQueue \ which support the bulk of data exchange between processes and the \verb PclPersistentQueue , which provides access to permanent data storage for input and output, and a disk-based, high-latency communication mode.\\
To assist the usage of multiple queues inside the framework, a generic queue array class is provided, called \verb PclQueues. To obtain the final objects used inside the framework, the \verb PclQueues \ is instantiated with a single queue object as template parameter, as in the following \verb typedefs :
\begin{verbatim}
PclQueues< PclQueue<AnyData> >  FrameQueues;
PclQueues< PclXQueue<AnyData> > FrameXQueues;
\end{verbatim}
where the CORBA container AnyData is actually a frame object.\\
The inheritance relationship between the queue classes is depicted in UML language in Fig. \ref{queuefig}.\\
The \verb PclQueue \ is a plain templatized queue. It provides overload of operators \verb << \ and \verb >> \ as aliases to \verb put \ and \verb get \ operations. A noteworthy extension is the addition of a \verb spy \ method, to get a copy of the next item contained in the queue. This is mainly used for real-time sampling of the flowing data.\\
The \verb PclXQueue \ purpose is to distribute its items to multiple clients. To meet the requirement, this queue is assisted by a circular buffer of indexes with a given maximum size. It can supply the same items to different clients. To transparently provide the queue behavior, the last element index is kept by the client side interface and submitted alongside with the request for the next item. This type of queue is used for buffering data to be processed for monitoring purposes, and log messages.\\
All the queue classes produce two warning triggers, \verb waterMark \ and \verb nearFull , based on a threshold on the number of waiting items in the queue.\\
The \verb PclPersistentQueue \ was first designed as an high latency, disk based queue. Providing input and output functionality alone was then a simple extension.\\
This queue turns out to be the main I/O interface to disk data for acquisition and analysis as well. The queue abstraction for fetching and writing data fits to the access pattern typical for AURIGA analysis.\\
The \verb PclPersistentQueue \ uses the frame library \cite{framelib} for data file access. Hence, it also acts as a wrapper to this library, and is the usual interface for data I/O.\\
Internally, the queue relies on a linked list of frame files \cite{framelib}, each node of which bears the file name, the begin and end times. The list is kept ordered by increasing begin time. Two pointers to this list are maintained, one for the current input file and one for the output. To select the behavior an instance of the queue (Full, Input, Output) it is thus sufficient to properly enable/disable the usage of one of the two pointers. The file list can be built offline at instantiation time (for Input only), or dynamically in Full queue mode.\\
Methods to determine and set input and output boundaries are provided, using the frame begin time as a key. This is the most natural key for humans to specify a dataset interval. The granularity at which this specification works is that of a frame, which for the AURIGA data is about $3$ seconds.\\
An useful feature for the acquisition is the ability of this queue to switch on-the-fly the disk it writes to. This allows the acquisition process to continue without interruption over the operation of disk changing.\\
A derived queue, the \verb PclMirroringQueue , is conceived to write the same data simultaneously on two different disk, in order to overcome the eventuality of a faulty disk.\\
The persistent queue hierarchy is presented in figure \ref{persqueuefig}.

\subsection{\label{corba}CORBA interface}
Every interprocess communication is handled by a unified CORBA interface. 
The strong type checking allowed by CORBA at the remote interface definition level makes the interface itself semantically meaningful.\\
Another interesting aspect of the CORBA-C++ couple is exception handling: the possibility of using exceptions over remote method invocation allows to design a more sleek and elegant interface.\\
The CORBA interface defined for the PCL framework includes:
\begin{itemize}
\item Data exchange: \verb GetData , \verb SpyData ;
\item Monitoring and configuration info: \verb GetConfig , \verb GetInfo , \verb SendInfo ;
\item Control interface: \verb Boot , \verb ShutDown , \verb Config , \verb Abort , \verb Start , \verb Stop , \verb Reset ;
\item Status and logging info: \verb GetLogMessage , \verb GetFsmStatus , \verb GetQueueStatus .
\end{itemize}
The data communication method provided by this interface is a smart polling. Clients try to get new data every few seconds. At each attempt, all the data currently present in the queue are fetched. This avoids the necessity of tuning the polling interval to keep up with the acquisition speed (see ref. \cite{tesi}).

\subsection{\label{control}Control}
\begin{figure}
\includegraphics[width=80mm]{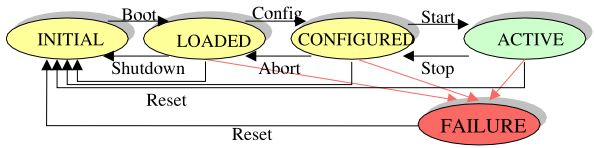}
\caption{Process control FSM.} \label{controlfig}
\end{figure}
At any time, the current state of a PCL process is determined by the Finite State Machine (FSM) depicted in fig. \ref{controlfig}.\\
Control commands are translated to transitions on this FSM.\\
As it is launched, a process goes to \verb INITIAL \ state. At this point, only minimal, non resettable initialization code is executed, and usually it is framework code.\\
The next state, \verb LOADED, implies that the configuration independent part of the initialization has been executed. During this phase, the configuration file for the process is read and parsed - but not used yet.\\
Then, in the \verb CONFIGURED \ state, the configuration information is used to complete the initialization. At this point, the process is ready to run its main job.\\
In fact, the only action performed by the transition to the next state, \verb ACTIVE , is the activation of the working thread.\\
A special transition alias, \verb Reset , brings the process state back to \verb INITIAL \ by executing in turn all the proper transition, depending on the current state, in order to ensure state consistency.\\
In addition to the regular states, there are two special ones: \verb FAILURE \ indicates an unrecoverable error, usually arisen during a state transition. The only action possible from this state is a \verb Reset . In this case, the \verb Reset \ special transition assumes the last regular state as the current one to determine the transitions to trigger.\\
The other special state is \verb INTRANSITION : it is set during any transition, and properly speaking it is not a state, but rather a flag that inhibits any transition command while set.

\subsection{\label{template}Template agent}
\begin{figure}
\includegraphics[width=80mm]{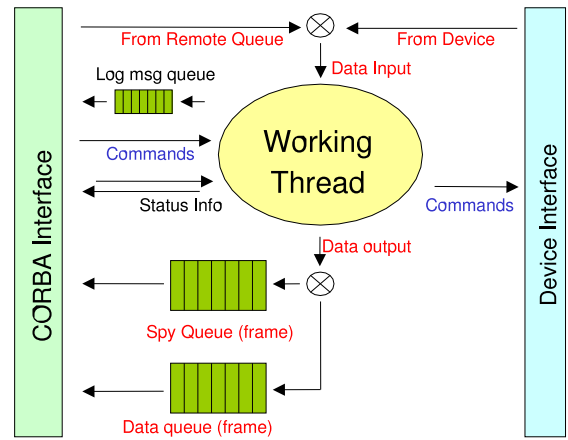}
\caption{Template process architecture.} \label{processfig}
\end{figure}
The PCL framework provides a template for the agents to accommodate their custom code. In an object oriented design, this pattern is best implemented by means of a base class bearing virtual methods.\\
The base class is \verb PclProducer , which provides a generic implementation for most of the virtual methods. The method that actual agents must override is the \verb Work \ method. The semantics of the \verb Work \ method is as follows: the method is cyclically called inside a dedicated thread continuously when a process is in \verb ACTIVE \ state.\\
The PCL framework provides to the specialized agents some virtual methods, called hooks, bound to every state transition defined in the control FSM. This gives the agent a structured way to define the initialization, configuration, and reset code.\\
More hooks are provided to allow the agents the finest customization ability to many operations: data communication, disk access, configuration. For most agents, though, the generic code supplied by the PCL is sufficient.\\
From the client side, PCL remote methods are accessed conveniently through the \verb PclClient \ interface. This interface presents an essential interface to all the methods, taking care transparently of most of the communication details, including type conversion to and from CORBA types, and indexes management.

\subsection{Configuration}
A generic configuration system is supplied by the PCL.\\
The configuration is meant to be edited at run-time, by means of the GUI or the command line tool, and saved to disk to be picked up automatically at the next system startup. Some configuration data is updated dynamically, such as the latest run number. This information has to be automatically saved to disk to be available for the next restart.\\
To address the aforementioned issues and to maintain consistency in persistent data format, the \em frame \em \cite{framelib} format was chosen, and a mechanism to store arbitrary data structures inside frames was developed.\\
The class \verb PclStructBase \ allows to serialize any data structure contained in an object derived from it. Data members are serialized transparently, while some help from the programmer is required to handle referenced dynamic structures, i.e. explicitly notifying the class the location and size of the attached item.\\
The array containing the serialized class is then stored in a section of the frame reserved for user data. Configuration frames are saved and read back automatically. File names are encoded by using the processes name.\\
Configuration frames are also used at runtime to allow clients, like the GUI, to request or set a process configuration. The frames are transparently sent through the CORBA interface. This is the main advantage of resorting to the frame format for storing configuration information: to transparently use the existing communication infrastructure.

\section{\label{agents}DAQ AGENTS}
The data acquisition agents contain the actual functionality of the acquisition system. The agents are PCL processes, and are labeled by an internal unique name of three letters.\\
Acquisition agents can be classified according to the schematic in \ref{processfig}: a front-end, composed by ANT, GPS, AUX; a collector layer, composed by BLD and LOG; storage, represented by DSK. Non-PCL processes are the clients, MON, GUI, DSC.\\

\subsection{Antenna - ANT}
ANT is responsible for acquisition of data buffers from the main HPE 1430A ADC internal memory, through the VXI, MXI, PCI hardware interfaces. At lowest level, bus handling is provided by National Instruments NI-VXI drivers.\\
The programming interface of this driver provides blocking functions. To keep the process responsive to external commands, all the driver related commands must be issued from a thread separated by the communication ones. The standard working thread provided by the PCL perfectly fits to the task.\\
The ADC accuracy being limited to 23 bits, ANT converts the ADC readout data to 32 bit floating point numbers in IEEE 754 format (\em float\em). Since this format assigns 24 bits to the mantissa, there is no loss of precision; meanwhile, data are multiplied by the ADC calibration constant, derived from its configuration register so that the sample values give directly the ADC input voltage.\\
This signal data is stored in frames, together with information about the sampling setup, ADC configuration, run number. The initial timestamp can not be assigned at this point: it will be obtained by the GPS process and inserted by the BLD.
\subsection{Auxiliaries - AUX}
AUX is similar to ANT, the main difference being the source ADC, the HPE 1413A. Unlike the single, high quality antenna channel, auxiliary signal spans multiple, low rate channels.\\
The different auxiliary channels are multiplexed in one single stream of readings, so that it is demanded to the software the task of separating the channels and setting the timestamps. This can not be done inside the AUX agent, because to separate the data into different frames the initial timestamps have to be known. This duty is then demanded to the BLD process.

\subsection{GPS}
GPS acquires initial timestamps of data buffers produced by the ADCs, converts them in a suitable format for later processing, and makes them available to the next agent in the acquisition chain.\\
Each timestamp is automatically generated by the custom GPS receiver, as explained in \ref{hardware}, and sent to a RS232 serial interface. The main thread of the GPS agent spends most of its time waiting for new timestamps to be available from the serial interface.\\
The alphanumeric timestamp supplied by the GPS is then converted to the PCL time type, implemented by two 32 bit integers, one representing seconds in Unix time convention, and one holding nanoseconds.\\
Converted timestamp are then put in separate queues, depending on the ADC they are related to. The system can discriminate different ADCs by looking at the interrupt line the trigger came from.

\subsection{Frame Builder - BLD}
BLD assembles partial data descriptions coming from the front-end agents to produce complete data frames. It is hence responsible to guarantee consistence of data, by implementing an array of checks, and to cope whenever possible with temporary failures of the auxiliary ADC or GPS receiver. The acquisition system always assumes the main antenna ADC up and working, when acquiring data.\\
Data blocks coming from the ADCs are fetched by using the standard PCL client interface, and are bound to the corresponding timestamps coming from the GPS agent. Consistency checks are performed for each frame to ensure that no time leaps or spurious timestamps are present, and eventually try to repair this failures by interpolating over the latest known good values.\\
A further time conversion is performed inside BLD: the time origin is set 10 years later, and yearly leap seconds are not taken into account, as requested by the Frame \cite{framelib} format.\\
A special treatment is required by auxiliary data: they have to be demultiplexed, and initial timestamps have to be assigned. Since the timestamps produced by the GPS are bound to the ADC multiplexed frames, the correct frame timestamps must be calculated by interpolation.\\
Antenna and auxiliary frames are then stored in two separate PCL queues, to make them available to next agents and to the real-time data monitoring.\\
Finally, the BLD takes care of setting for each frame the user part of the data quality. This is set through the GUI by the acquisition operators, e.g. when the detector is under maintenance.

\subsection{Disk - DSK}
DSK is responsible of the persistent storage of complete data frames produced by the BDL to disk.\\
This apparently trivial duty is complicated by the fact that disk access is often one major cause of environment change for application software, and consequently one major cause of failure. Some examples of such dangerous environmental changes are full disks, dramatical slowdowns due to concurrent activity, filesystem corruption, hardware failure.\\
To address this range of problems, DSK uses a special persistent queue, provided by the PCL, called \verb PclMirroringQueue . This queue takes care of concurrently writing the same data frames to two different disks, thus giving substantial fault tolerance.\\
Additionally, an alternative pair of disks can be configured, in order to automatically switch to write the data files on them whenever the other ones are full, or the operator wishes to do any maintenance on them. In this configuration, DSK handles four different disks.

\subsection{Logger - LOG}
Any PCL process produces log messages by simply invoking the static method \verb notify \ provided by the PCL. Log messages are encoded in order to be stored in a queue and sent over through the CORBA interface. Each PCL process has a dedicated queue for log messages.\\
LOG periodically polls all the active PCL processes, including itself, to collect the log messages waiting in their queues. Gathered messages are then stored in a special queue, of the same type as the monitoring queues. This allows LOG to provide the message history and updates to any number of clients. The use case that exploits this data structure, is the handling of multiple GUIs, monitoring the system at the same time.

\subsection{User interface - GUI,DSC}
\begin{figure}
\includegraphics[width=40mm]{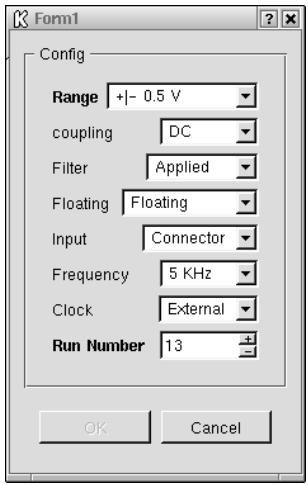}
\caption{Configuration panel for the ANT process.} \label{antconffig}
\end{figure}

GUI and DSC are not PCL processes. They only use the client side interface provided by the PCL.\\
\begin{figure}
\includegraphics[width=75mm]{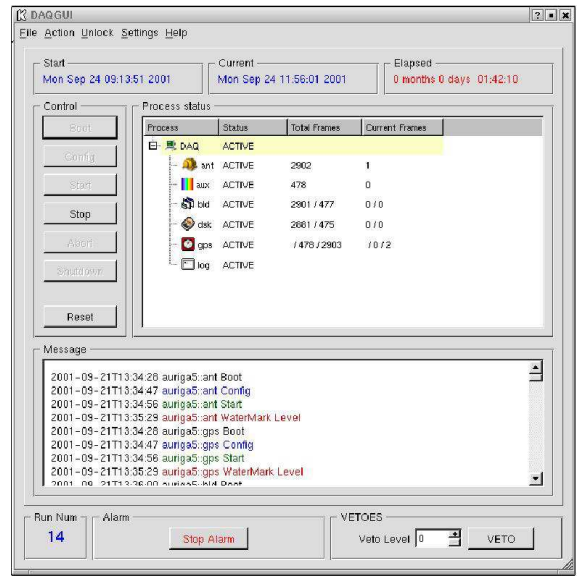}
\caption{The main GUI interface.} \label{guifig}
\end{figure}
A picture of the GUI main screen is reported in fig. \ref{guifig}.\\
The main component is the process window. It is updated at regular, user definable, intervals. It displays all the active PCL processes (under the given CORBA name server), and for each one of them, its status, and some statistics about its data queues. This information is sufficient to spot and identify most failures of the acquisition system.\\
At the left of the process window there are the control buttons: they act on the process status by triggering transitions on the control finite state machine. The pseudo-process DAQ applies the required transition to all the processes; when a process receives a request not applicable to its current status, it simply ignores it.\\
The process window and command buttons are able to handle automatically new PCL processes, without any need to update the code.\\
The remaining information provided by the GUI relies on specific processes. Log messages are fetched from the LOG process, and it is up to the GUI to associate a different color to messages of different type, and to bind execution of actions, typically alarms, to some log message codes. Alarms are also triggered when any data queue size exceeds a threshold, as this usually means that the next process in the acquisition chain is stuck.\\
The current run number is provided by the ANT process, while the timing information and current data quality setting come from the BLD. Quality setting can be modified from the GUI, and the BLD is automatically notified of the change.\\
Custom configuration panels, specialized for each process, are invoked by double-clicking a process from the list. An example configuration panel, the one for ANT, is displayed in figure \ref{antconffig}. Configuration update is only allowed when the process is in \verb LOADED \ state. Custom information panels are raised by right-clicking a process.\\
%%GUI impl details? and pic?
When started, the GUI disables any command option, and just does monitoring. There is a specific menu option to unlock the GUI and enable the command features.

\subsection{Monitor - MON,AIM}
MON is a simple acquisition monitoring tool. It is able to display in real-time data frames from any PCL process.\\
\begin{figure}
\includegraphics[width=50mm]{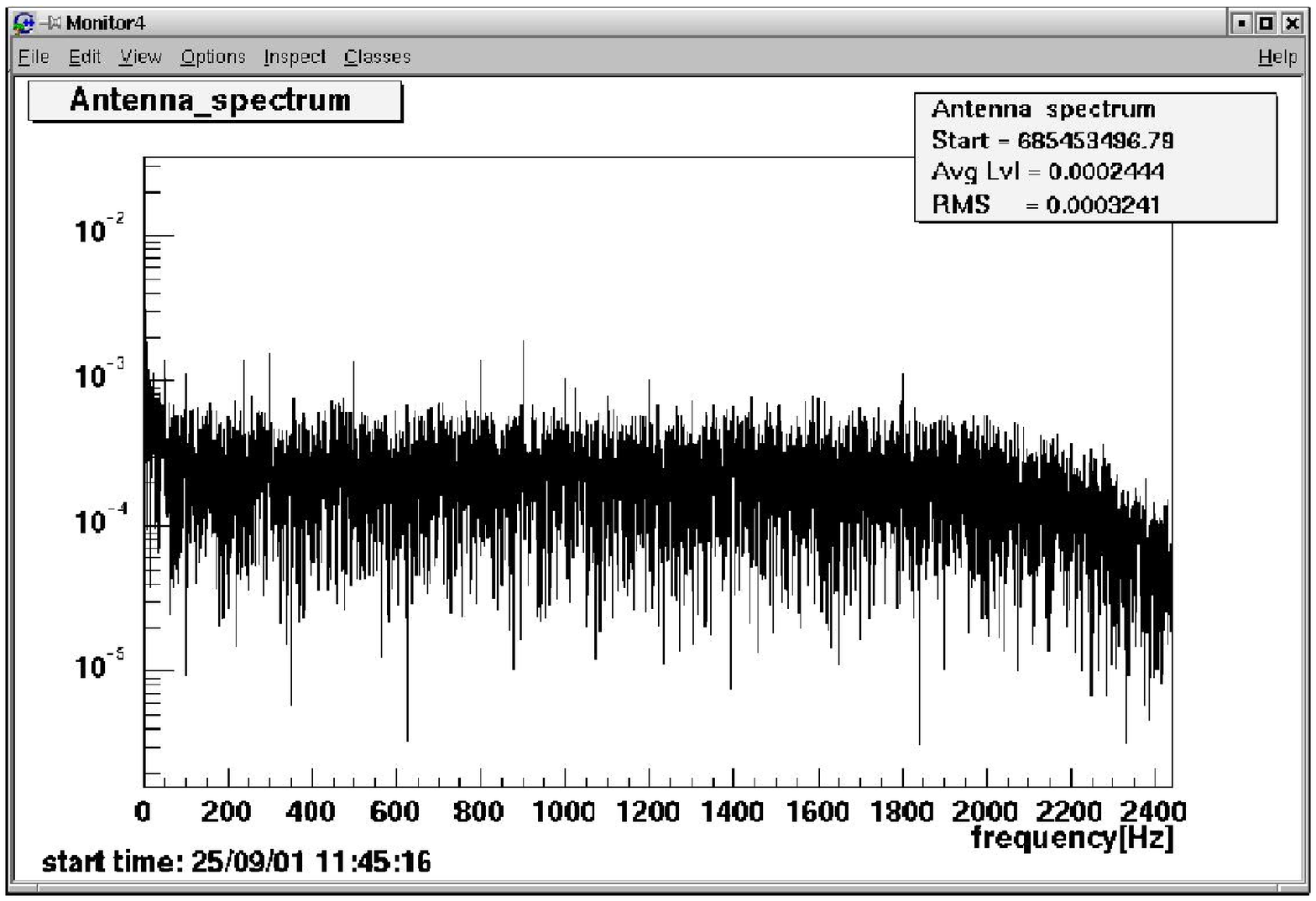}
\includegraphics[width=30mm]{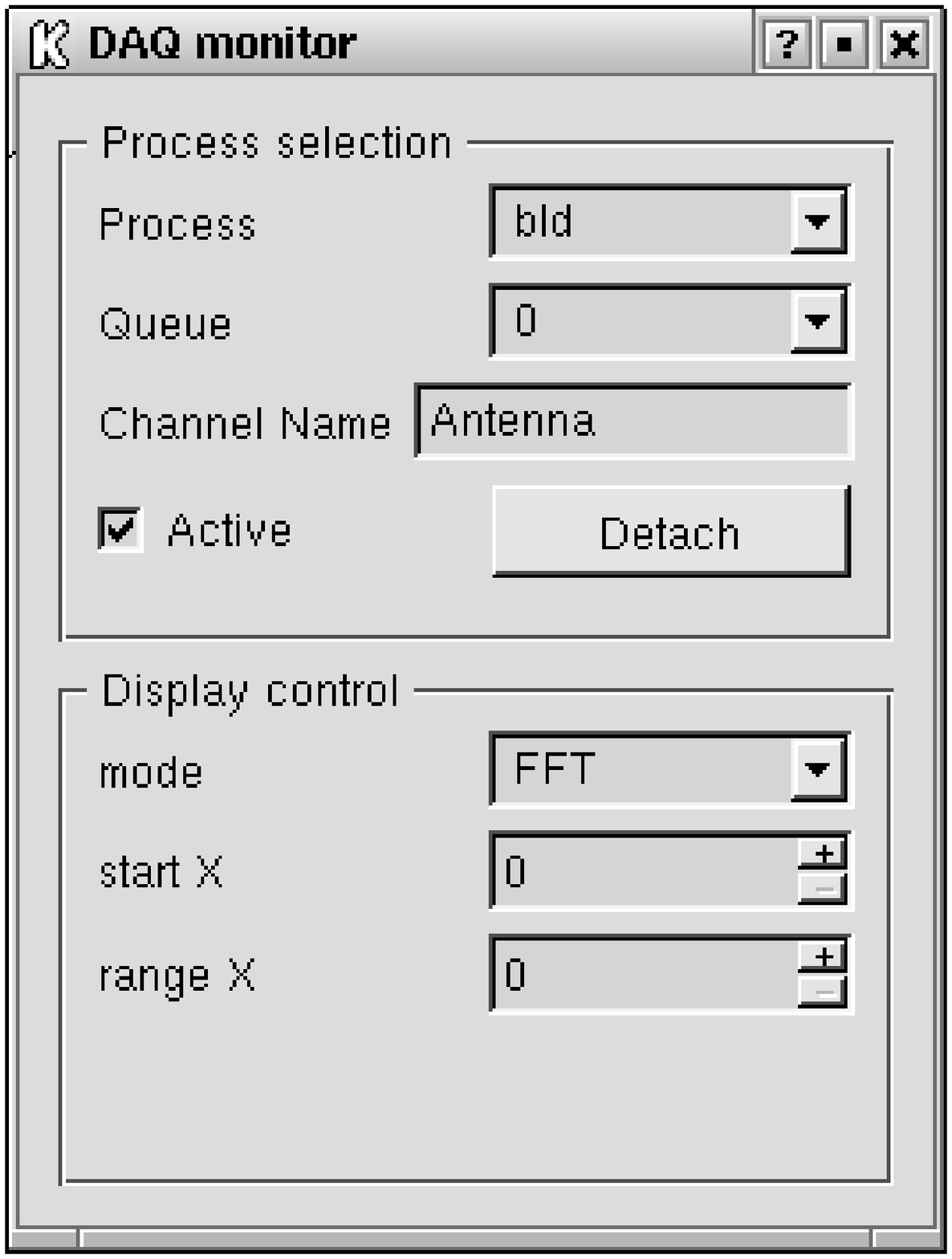}
\caption{The MON user interface, displaying a spectrum of the unconnected ADC input noise.} \label{monfig}
\end{figure}
The simple graphical user interface, displayed in fig. \ref{monfig}, allows to select a data channel by name, to show the signal in the frequency domain by discrete fourier transform, to restrict the display to a given time or frequency range.\\
The visualization interface is based on the VEGA framework \cite{vega}, derived from ROOT \cite{root}.\\
The \verb Detach \ button actually interrupts the monitor and gives control to the underlying ROOT instance. This allows to directly use the graphical analysis methods provided by the framework.\\
An alternative interface to the monitoring application, called AIM, is based on the ROOT command line interpreter, rather than a GUI. AIM extends the functionality by allowing to work on data frames coming from different sources, such as files stored on disk.

\section{Conclusions}

\subsection{Software design}
The main features of the system design were: fully Object Oriented design methodology, and extensive use of Open Source components and development tools.\\
Quality measures of an OO design are the stability of object interfaces, and the extent of components reuse. In both respects, the acquisition system performs well.\\
The programming interface provided by the PCL framework remained quite stable since the beginning of application software development, i.e., the specific agents code. As the specific process code was developed, the PCL was modified mostly to add new functionality, rather than to revise existing object interfaces.\\
Experience with Open Source software was commented thoroughly in a dedicated paragraph. The most valuable technical advantage of using software licensed as Open Source has been for us long-term maintenance. Some software components used along the development were changed, other ones underwent radical management changes. From our user point of view, this never created policy or compatibility problems.\\
The programming technique of wrapping whenever possible the interface of external software components inside custom adapter layers yielded better design freedom and ease of maintenance.

\subsection{Performance}
The throughput requirements of the AURIGA experiment are not very high. The pure data throughput is about 30 $kB/sec$. Using modern machinery, the performance of the system exceeds by more than two orders of magnitude the requirement. Some throughput benchmarks are reported in the following table:\\
\\
\begin{tabular}{rrr|rrr}
\multicolumn{3}{c}{dual Xeon@2400, networked} & \multicolumn{3}{c}{dual PIII@800, local}\\
Block & Blk/sec & $kB$/sec & Block & Blk/sec & $kB$/sec\\
16 & 50 & 800   & 16 & 49 & 784\\
64 & 50 & 3200  & 64 & 35 & 2240\\
192 & 50 & 9600 & 160 & 27 & 4320\\
\end{tabular}
\\

A more sensible requirement for the acquisition system was reliability. To attain this objective, the multi-agent architecture proved itself very effective. The capability of restarting individual agents without disrupting the system restricts the single point of failure to the agent dealing with the main antenna ADC. This restriction could eventually be overcome only by using two redundant front-end ADCs.\\
The system showed no unexpected crashes or behavior during 1 year long continuous usage in the AURIGA test facility, and is ready for its duty in the upcoming new acquisition run of the AURIGA experiment.

%%\section{-- STUFF --}

\begin{acknowledgments}
It is a pleasure to acknowledge the AURIGA team for
providing us continuous feedback and suggestion on
the data acquisition system. One of us (A. C.)
would also thank the Laboratori Nazionali di Legnaro for
the kind hospitality during his graduating thesis on computer science.
\end{acknowledgments}


\begin{thebibliography}{9}   % Use for  1-9  references
%\begin{thebibliography}{99} % Use for 10-99 references

%\bibitem{accelconf-ref}
%http://www.cern.ch/accelconf

\bibitem{tesi}A. Ceseracciu, \em Progettazione e realizzazione di un sistema di acquisizione dati per l'esperimento AURIGA \em (italian), Dipartimento di Elettronica e Informatica, Universit\'{a} di Padova, Padova (2001)
\bibitem{gpstime}M. Cerdonio et al., \em Sub-Millisecond Absolute Timing: Toward an Actual Gravitational Observatory\em, Mod. Phys. Lett A 12 (1997) 2261
\bibitem{olddaq}A. Ortolan et al. in Proc. of the 2nd E. Amaldi Conference on
Gravitational Waves, E. Coccia, G. Veneziano and G. Pizzella Eds.
(Word Scientific Singapore 1998) p. 204
\bibitem{omnidoc}Sai-Lai-Lo, D. Riddoch, D. Grisby, \em The omniORB version 3.0
User's Guide\em, AT\&T Laboratories Cambridge (2000)
\bibitem{omniweb}http://www.uk.research.att.com/omniORB
\bibitem{corba}http://www.corba.org
\bibitem{kdev}http://www.kdevelop.org
\bibitem{vega}http://wwwlapp.in2p3.fr/virgo/vega
\bibitem{root}http://root.cern.ch
\bibitem{virgo}http://www.virgo.infn.it
\bibitem{ligo}http://www.ligo.caltech.edu
\bibitem{framelib}B. Mours, \em A common data format for gravitational waves interferometers, Gravitational Wave Detection\em, Ed. by Tsubono, Fujimoto, Kuroda, Universal Academy Press Inc., Tokyo (1997) 27-30

%%-> Franz paper here
%% ... what else? ... maybe pete's paper ...?
%A.N. Other, ``A Very Interesting Paper'', EPAC'96, Sitges, June 1996.

%\bibitem{templates-ref}
%http://www.cern.ch/accelconf/templates.html

\end{thebibliography}

\end{document}